\documentclass[prd,twocolumn,showpcs,amsmath,amssymb,nofootinbib,preprintnumbers,balancelastpage]{revtex4}
\usepackage{epsfig}
\usepackage{amsmath}
\usepackage{bm}
\usepackage{times}
\usepackage{graphicx}
\usepackage{color}
\usepackage{slashed}
\usepackage{graphicx}
\usepackage{amsmath, amssymb}

\def\bea{\begin{eqnarray}}
\def\eea{\end{eqnarray}}
\def\ba{\begin{eqnarray}}
\def\ea{\end{eqnarray}}
\def\be{\begin{equation}}
\def\ee{\end{equation}}

\begin{document}
\preprint{Submitted to JHEP}

\title{Finite Temperature Structure of the Compactified Standard Model}

\author{Jonathan M. Arnold, Bartosz Fornal and Koji Ishiwata\\
\textit{California Institute of Technology, Pasadena, CA 91125, USA}\\
}
\date{\today}

\begin{abstract}
  We analyze the finite temperature structure of the standard model
  coupled to gravity with one and two dimensions compactified (on a
  circle and a torus).
  We find that finite temperature effects wash out any vacua
  which exist at zero temperature.  We also discuss the possibility
  of transitions between the four-dimensional universe and the
  lower-dimensional spacetimes.
\end{abstract}

\maketitle
\bigskip

\section{Introduction}
It has recently been demonstrated that the standard model coupled to
gravity, besides the unique four-dimensional vacuum, may also have
vacua stabilized by Casimir energies of standard model particles if
one spatial dimension is compactified on a circle \cite{Arkani} or if
two spatial dimensions are compactified on a torus \cite{AFW}.  The
vacuum of the low-energy effective theory for the one-dimensional compactification can be either anti de
Sitter, Minkowski, or de Sitter, whereas for the two-dimensional
compactification it is necessarily anti de Sitter. Such vacua
exist for a wide range of neutrino masses.

In this paper we extend the results of \cite{Arkani} and \cite{AFW}
and explore the stabilization of compact dimensions at finite
temperature. We calculate the free energy and use it to derive
formulas for the energy density and pressure. We then use Einstein's
equations to analyze the stability of the compact space. We find
that increasing temperature washes out any existing stable points.

At a given temperature, only particles of masses smaller than this
temperature have a considerable finite temperature effect on Einstein's equations.
This fact, along with the known standard model spectrum, allows us to
analyze the stability of compact dimensions up to temperatures around
$100\ \rm{GeV} \simeq 10^{15} \ \rm{K}$.

Finally, we explore the possibility of our universe spontaneously
compactifying to a lower-dimensional spacetime.  The tunneling rate
for producing a bubble of such a compact geometry can be estimated by
calculating the Coleman-de Luccia-like instanton \cite{Sean}.  Not
surprisingly, we find that the rate for such a process is exceedingly small.


\section{Compactified spacetime at finite temperature}
In this section, we discuss the stability of compactified spacetimes
for nonzero temperatures.  The action of the standard model coupled to
gravity is given by,
\begin{eqnarray}\label{action}
S = \int d^4x \sqrt{-g_{(4)}}\left(\frac{1}{2}M_{\rm pl}^2
{\cal R}+{\cal L}_{\rm field}\right),
\end{eqnarray}
where $g_{(4)}$, $M_{\rm pl}\simeq 2.4 \times 10^{18}~{\rm GeV}$,
$\mathcal{R}$, and ${\cal L}_{\rm field}$ are the determinant of the
4D metric, Planck mass, Ricci scalar, and standard model Lagrangian,
respectively. We implicitly include the 4D cosmological
constant in $\cal{L}_{\rm field}$. Variation of this action with respect to the
metric $g_{\mu\nu}$ yields Einstein's equations,
\begin{eqnarray}
G_{\mu\nu}= \frac{1}{M_{\rm pl}^2} T_{\mu \nu},
\label{eq:EinsteinEq}
\end{eqnarray}
where $G_{\mu\nu}$ is the Einstein tensor and $T_{\mu\nu}$ is the
stress-energy tensor containing the cosmological constant term.
Starting from these Einstein's equations, we investigate
the structure of the standard model for one- and two-dimensional
compactifications.

\subsection{One-dimensional compactification}
\begin{figure*}[t]
\centerline{\scalebox{1.00}{\includegraphics{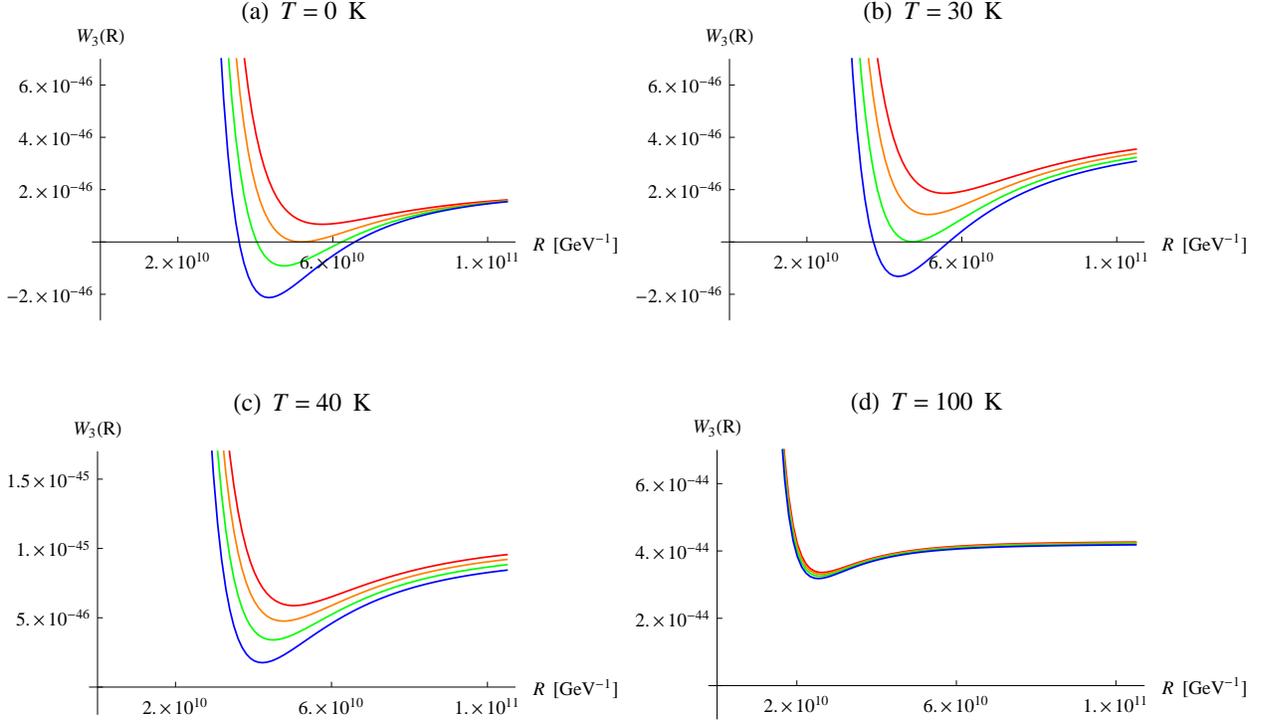}}}
\caption{\footnotesize{The RHS of equation (\ref{box_eq}), $W_3(R)$, as a
    function of the compactification radius $R$ for normal hierarchy
    Dirac neutrinos with the lightest neutrino mass of $6\times
    10^{-12} {\rm \ GeV}$ (red), $7\times 10^{-12} {\rm \
      GeV}$ (orange), $8\times 10^{-12} {\rm \ GeV}$
    (green), and $9\times 10^{-12} {\rm \ GeV}$ (blue) for
    temperatures: (a) $T = 0$, (b) $T = 30 \ \rm K$, (c) $T = 40 \ \rm
    K$, and (d) $T = 100 \ \rm K$, respectively. Note that the ranges
    in figures (c) and (d) are larger than for the first
    two plots.  $M_{\rm pl}$ has been set to 1 for simplicity.}}
\end{figure*}
Here we discuss the finite temperature effects for the case of one
dimension compactified on a circle.  Working in four dimensions, we
consider the spacetime interval,
\begin{eqnarray}
ds^2 = g_{(3)\mu\nu}\, d x^\mu d x^\nu + R(t)^2\, d \phi^2\ ,
\end{eqnarray}
where $g_{(3)\mu\nu}$ is the metric on the noncompact 3D spacetime
with $\mu, \nu = 0, 1, 2$, and the compact coordinate $\phi \in [0,
2\pi)$.  We assume a homogeneous and isotropic 3D metric,
\begin{eqnarray}
g_{(3)\mu\nu} = \left(
                  \begin{array}{ccc}
                    -1 & 0 & 0 \\
                    0 & \frac{a(t)^2}{1-\kappa \,r^2} &  0 \\
                    0 & 0 & a(t)^2\,r^2 \\
                  \end{array}
                \right)\ ,
\end{eqnarray}
where $a(t)$ is the scale factor of the noncompact
space, and $\kappa=1,0, \mbox{\rm or} -1$ corresponds to a closed, flat, or open metric,
respectively. We choose a stress-energy tensor consistent with the
symmetries of the metric,
\begin{eqnarray}
T^{\mu}_{\ \ \nu} = \left(
                  \begin{array}{cccc}
                    -\rho_{3} & 0 & 0 & 0 \\
                    0 & p_{3} & 0 &  0 \\
                    0 & 0 & p_{3} & 0  \\
                    0 & 0 & 0 & p_R \\
                  \end{array}
                \right)\ ,
\end{eqnarray}
where $\rho_{3}$ is the energy density, and $p_{3}, p_R$ are the
pressures in the noncompact and compact space, respectively, all being
functions of $R$ and $\beta = 1/(k_B T)$, with $T$ -- temperature.
Einstein's equations (\ref{eq:EinsteinEq}) take the form,
\begin{eqnarray}
&&2\,\frac{\dot{a}}{a}\frac{\dot{R}}{R}+\frac{\dot{a}^2}{a^2}
+\frac{\kappa}{a^2} = \frac{\rho_{3}}{M_{\rm pl}^2}\ ,
\label{eq:3D1} \\
&&\frac{\ddot{R}}{R}+\frac{\dot{a}}{a}\frac{\dot{R}}{R}+ \frac{\ddot{a}}{a}
=  -\frac{p_{3}}{M_{\rm pl}^2}\ ,
\label{eq:3D2}\\
&&2\,\frac{\ddot{a}}{a}+ \frac{\dot{a}^2}{a^2} +\frac{\kappa}{a^2}
=  -\frac{p_R}{M_{\rm pl}^2}\ .
\label{eq:3D3}
\end{eqnarray}
Here the dot denotes the derivative with respect to time.  This yields
the equation of motion for $R$,
\begin{eqnarray}\label{box_eq}
  \frac{\ddot{R}}{R}+2\frac{\dot{a}}{a}\frac{\dot{R}}{R}
  =  \frac{1}{M_{\rm pl}^2} \left(- p_{3} +\frac{1}{2}\,\rho_{3}
+ \frac{1}{2}\,p_R \right) \equiv W_3(R)\ .
\end{eqnarray}
The left hand side of (\ref{box_eq}) can be written as $-\square \log R$,
where the d'Alembertian $\square \equiv
D_{\mu}D^{\mu}$. The energy density and
pressures are calculated from the free energy using standard
thermodynamic relations \cite{Kolb},
\begin{eqnarray}
\rho_{3} &=& \frac{1}{2\pi R\, V_2}\frac{\partial \left(\beta F_3\right)}
{\partial \beta}\bigg|_{a, R}\ ,
\label{rho} \\
p_{3} &=& -\frac{1}{2\pi R}\frac{\partial F_3}
{\partial V_2}\bigg|_{R, \beta}\ ,
\\
p_R &=& -\frac{1}{2\pi V_2}\frac{\partial F_3}{\partial R}\bigg|_{a, \beta}\ ,
\label{pressure}
\end{eqnarray}
where $V_2$  is the volume of the noncompact space.  The
total free energy is obtained by summing the contributions from all
standard model particles\footnote{One has to be careful when including
  contributions from quarks and gluons. At temperatures of the order
  of the meson masses, for example, meson fields must be treated as
  fundamental.}  and the cosmological constant term. It is given by,
\begin{eqnarray}\label{free_all}
F_3 =
\sum_{\rm particles} F^{b/f}_3 +2\pi R\,V_2\, \Lambda\ ,
\end{eqnarray}
where $\Lambda$ is the cosmological constant, and $F^{b/f}_3$ is the free energy for a
particle of mass $m$. It is obtained by following the
steps outlined in \cite{Kapusta},
\begin{eqnarray}\label{free}
F^{b/f}_3&\!=\!&
V_2\,(\pm \,d) \int \frac{d^{2} k}{(2\pi)^{2}}\sum_{n
 = -\infty}^\infty\bigg[\,\frac{1}{2} \sqrt{\vec{k}^2 +m^2+\tfrac{n^2}{R^2}}
\nonumber\\
&&\ \ \ \ \ \ \ \ \ \ \ \ \ \ \ \ \ \ \ + \,\frac{1}{\beta}\log\left(1\mp e^{-\beta\sqrt{\vec{k}^2 +
      m^2+\frac{n^2}{R^2}}}\right)\bigg]
\nonumber \\
&\equiv& V_2\left( \rho_{\rm Cas(3)}^{b/f}+f^{b/f}_{(3)}\right),
\end{eqnarray}
where the upper sign corresponds to bosons, the lower one to fermions, and
$d$ is the number of degrees of freedom. The quantity
$V_2 \rho_{\rm Cas(3)}^{b/f}$ is the zero temperature Casimir energy
and $V_2 f^{b/f}_{(3)}$ is the finite temperature contribution to
free energy.\footnote{Note that while the energy
density $\rho_3$ and pressure $p_3$ are mass dimension four,
the Casimir energy density $\rho_{\rm Cas(3)}^{b/f}$
and free energy density $f^{b/f}_{(3)}$ are mass dimension three.}

It is straightforward to show that formula (\ref{free_all}) is free of
divergences.  The 3D Casimir energy density for a particle of mass $m$ is given by,
\bea\label{15}
&&\!\!\!\!\!\!\!\rho_{\rm Cas(3)}^{b/f} =
\mp d\bigg[
\frac{\Gamma\left(-2\right)}{16\pi} m^4 R
+\frac{m^2}{4\pi^3}\frac{1}{R}\sum_{n=1}^\infty\frac{1}{n^2}K_2(2\pi
\,n \,m\,R)\bigg]
\nonumber \\
&&\ \ \ \ \  \ \ \ \equiv  \rho_{\rm Cas(3)div}^{b/f}+ \tilde{\rho}_{\rm Cas(3)}^{b/f}\ ,
\eea
where $K_n(x)$ is the modified Bessel function of the second kind, $V_2\rho_{\rm Cas(3)div}^{b/f}$
is the divergent part of the Casimir energy, and $V_2\tilde{\rho}_{\rm Cas(3)}^{b/f}$ is the
finite regularized contribution.  In
the massless limit formula (\ref{15}) reduces to,
\begin{eqnarray}\label{massless}
\rho_{\rm Cas(3)}^{b/f}\Big|_{m=0} = \mp\,
\frac{d}{720\, \pi R^3}\ .
\end{eqnarray}
Note here that,
\begin{eqnarray} \Lambda = \Lambda^{\rm obs} +
\Lambda^{\rm q. corr.}\ ,
\end{eqnarray}
where the superscript ``{\rm obs}'' indicates the observed value
(i.e., $\Lambda^{\rm obs}\simeq3.1\times 10^{-47}~{\rm GeV}^4$
\cite{pdg}), while ``{q.corr.}'' denotes the quantum correction in
flat space given by,
\begin{eqnarray}
  \Lambda^{\rm q.corr.} =
  \sum_{\rm particles}(\pm\, d) \,\frac{\Gamma\left(-2\right)}{32\,\pi^2} m^4\ .
\end{eqnarray}
Since $\rho_{\rm Cas(3)div}^{b/f} = -2\pi R\,\Lambda^{\rm q. corr.}$, all the divergences cancel and the free energy can be
expressed in terms of finite quantities using equation
(\ref{free_all}),
\begin{eqnarray}\label{free3D}
F_3/V_2&=&
\sum_{\rm particles}\left[ \tilde{\rho}^{b/f}_{\rm Cas(3)}+f^{b/f}_{(3)}  \right]
+2\pi R\, \Lambda^{\rm obs}
\nonumber \\
&\equiv&\tilde{\rho}_{{\rm Cas}\Lambda(3)} + f_{(3)}\ .
\end{eqnarray}
The first term in the second line of (\ref{free3D}) is the Casmir energy including the cosmological constant term
and has no temperature dependence. The second term is the finite temperature
contribution.

In summary, the dynamics of the field $R$ is governed by equation
(\ref{box_eq}), where $\rho_{3}, p_{3}, p_R$ are given by formulas
(\ref{rho})-(\ref{pressure}), with the free energy expressed through
equation (\ref{free3D}).  It is easy to check that equation
(\ref{box_eq}) at zero temperature becomes the equation of motion
resulting from the dimensionally reduced action, derived along the lines
of \cite{Arkani}, and  takes the form,
\begin{eqnarray}
\square \log R =\frac{R^3}{4\pi M^2_{\rm pl}}
\left(\frac{\tilde{\rho}_{{\rm Cas}\Lambda(3)}}{R^3}\right)' \ ,
\label{eq:3D_R_T0}
\end{eqnarray}
where the prime indicates the derivative with respect to $R$.

It has been shown \cite{Arkani} that depending on the neutrino masses,
their type (Dirac or Majorana), and choice of hierarchy, there might
exist a 3D anti de Sitter, Minkowski, or de Sitter vacuum of
the low-energy effective theory. Including
finite temperature effects has similar implications for all those
cases, therefore, for illustrative purposes, we discuss only the case
of Dirac neutrinos with normal hierarchy (figure 1).  Note that the
right hand side of equation (\ref{box_eq}), i.e., $W_3(R)$,
is analogous to the derivative of the potential, so it should be equal to zero at the
stationary points. In addition, $d W_3/d R < 0$ at the stationary point
implies stability, whereas $d W_3/d R >0$ corresponds to an
unstable stationary point.

At temperatures $T \ll m_e \simeq 6\times 10^{9} \rm \ K$ only the
photon, graviton, and neutrino contributions are relevant
in equation (\ref{free3D}). Figure 1 (a) shows the plot of $W_3(R)$
at $T=0$ for several lightest neutrino masses.
In the cases for which stationary points exist, the one appearing at smaller $R$ is stable, whereas
the one for larger $R$ is unstable. Those points match the ones
found along the lines of \cite{Arkani}.
We checked that for low
temperatures ($T\lesssim 10\rm \ K$) the finite
temperature effects are negligible. However, as soon as the temperature
reaches approximately 20 K, the curves slowly depart from
the zero temperature result and move upwards. Figure 1 (b) shows the plot
of $W_3(R)$ for $T=30 \rm \ K$.

With temperature further increasing, each curve moves up and eventually
fails to cross zero (figure 1 (c), (d)). At very
high temperatures, the contributions from other standard model particles
become relevant.  We have investigated the behavior of $W_3(R)$ up to a
temperature $T = 100\ \rm GeV \simeq 10^{15}\rm \ K$ and confirmed
that no zero points appear.

The behavior of $W_3(R)$ at high temperatures can be understood as follows. At a fixed
temperature, only particles with Kaluza-Klein mass
$m^2_{n}=m^2+n^2/R^2\lesssim T^2$ contribute to $f_{(3)}$, since the
particles with larger masses are Boltzmann suppressed. In addition,
particles with masses smaller than $T$ can be treated as radiation.
Then, the finite temperature contributions to $W_{3}(R)$ coming from
$\rho_{3}$ and $p_{3}$ (say $\rho^T_{3}$ and $p^T_{3}$, respectively)
cancel because $p^T_{3}\simeq\rho^T_{3}/2$ at leading
order.
Equation (\ref{box_eq}) now becomes,
\begin{eqnarray}
&&  \!\!\!\!\!\!\!\!\!\!\!\!\!\!\!\frac{\ddot{R}}{R}+2\frac{\dot{a}}{a}\frac{\dot{R}}{R}
\nonumber \\
 && \!\!\!\!\!\!\!\!\!\!\simeq  \frac{1}{2\pi R \,M_{\rm pl}^2}
\left(\tfrac{3}{2}\tilde{\rho}_{{\rm Cas}\Lambda (3)}
- \tfrac{1}{2}R\, \tilde{\rho}'_{{\rm Cas}\Lambda (3)}
- \tfrac{1}{2}R\, f^{T\prime}_{(3)}\right).
\label{eq:3D_Rlim}
\end{eqnarray}
The first two terms on the right hand side are the zero temperature contributions.
The third term is temperature dependent
and can be estimated as $f^{T\,\prime}_{(3)} \sim
-(TR)'T^3=-T^4$, where the factor $(TR)$ comes from summing over
Kaluza-Klein modes.
Therefore, in the high temperature limit, equation (\ref{box_eq}) takes the form,
\vspace{-2mm}
\begin{eqnarray}
\square \log R =\frac{R^3}{4\pi M^2_{\rm pl}}
\left(\frac{\tilde{\rho}_{{\rm Cas}\Lambda(3)}+\sigma_3 R\,T^4}{R^3}\right)',
\label{now}
\end{eqnarray}
where $\sigma_3$  encodes the number of standard model
particles whose mass is smaller than $T$.
For small $R$ we have $\tilde{\rho}_{{\rm Cas}\Lambda(3)} \sim 1/R^3$, so
the first term on the right hand side of equation (\ref{now}) is negligible for $T \gg 1/R$.
Since for all experimentally allowed neutrino masses the zero temperature
vacuum occurs at $R_{\rm vac} \approx 10^{10} {\ \rm GeV^{-1}}$,
the high temperature limit applies for $T \gg 1000 \ {\rm K}$.
This confirms that finite temperature effects remove any zero
points of $W_3(R)$ and, consequently, make the vacuum disappear.
It also explains why the universe would have expanded rather than  settle
in a three-dimensional vacuum existing for small $R$ at zero temperature.

\begin{figure*}[t]
\centerline{\scalebox{1.00}{\includegraphics{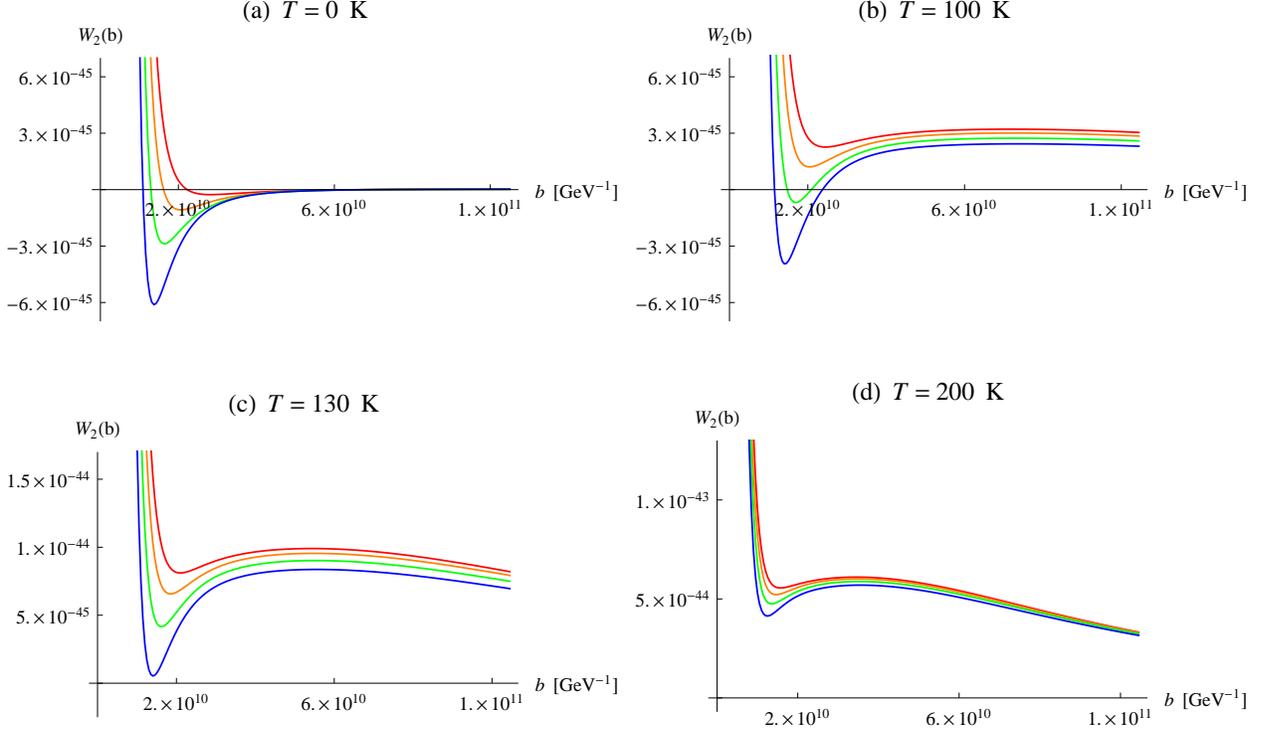}}}
\caption{\footnotesize{The RHS of equation (\ref{boxeq}), $W_2(b)$,
    as a function of the volume modulus $b$ for normal hierarchy Dirac
    neutrinos with the lightest neutrino mass of $1.0\times 10^{-11}
    {\rm \ GeV}$ (red), $1.5\times 10^{-11} {\rm \ GeV}$ (orange),
    $2.0\times 10^{-11} {\rm \ GeV}$ (green), and $2.5\times 10^{-11}
    {\rm \ GeV}$ (blue) for temperatures: (a) $T = 0$, (b) $T = 100 \
    \rm K$, (c) $T = 130 \ \rm K$, and (d) $T = 200 \ \rm K$,
    respectively. The ranges
    in figures (c) and (d) are larger than for the first
    two plots. $M_{\rm pl}$ has been set to 1 for simplicity.}}
\label{fig:t2plots}
\end{figure*}

\subsection{Two-dimensional compactification}
Next, we investigate the finite temperature structure of the standard
model for a toroidal compactification.  The four-dimensional spacetime
interval is given by,
\begin{eqnarray}
ds^2 = g_{(2)\mu\nu}(x) \,d x^\mu d x^\nu + t_{i j}(x) \,d y^i d y^j\ .
\end{eqnarray}
The indices $\mu, \nu$ are now $0$ or $1$, and,
\begin{eqnarray}\label{torus}
t_{i j} = \frac{b^2}{\tau_2}\left(
  \begin{array}{cc}
    1 & \tau_1 \\
    \tau_1 & |\tau|^2 \\
  \end{array}
\right),
\end{eqnarray}
where $b^2$ and $\tau = \tau_1 + i \tau_2$ are the volume and shape
moduli, respectively.  From here on we will assume $\tau = 1/2 + i
\sqrt{3}/2$ in all formulas for the reasons discussed in \cite{AFW}.
We also assume a homogeneous universe, thus $b$
depends only on time. For the noncompact spacetime, we parametrize
the line element as,
\begin{eqnarray}
ds_{(2)}^2 = -d t^2 + a(t)^2 d r^2.
\end{eqnarray}
The stress-energy tensor has the form,
\begin{align}
T^{\mu}_{~~\nu} =
\left(\begin{array}{cccc}
-\rho_{2} & 0 & 0 & 0 \\
0 & p_{2} & 0 & 0 \\
0 & 0 & p_b & 0 \\
0 & 0 & 0 & p_b
\end{array}\right),
\end{align}
where $\rho_{2}$, $p_{2}$, and $p_b$ are the energy density, pressure
in the noncompact space, and pressure in the compact space, respectively.

We can now proceed along the same lines as in the previous section. The 4D
free energy for a particle of mass $m$ is,
\begin{eqnarray}\label{casm}
  F^{b/f}_2
  &\!=\!&  V_{1}\,(\pm\, d) \int \frac{d k}{2\pi}\sum_{n_1, n_2
    = -\infty}^\infty\Bigg[\frac{1}{2}\sqrt{k^2 +m^2+t^{i j}n_i n_j}\nonumber\\
  &&\ \ \ \ \ \ \ \ \ \ \ \ \ \ \ \ \ \ \ \ \ \ +\frac{1}{\beta}\log\left(1\mp
    e^{-\beta\sqrt{k^2 + m^2+t^{i j} n_i n_j}}\right)\Bigg]
  \nonumber \\
  &\equiv&V_1\,\left(\tilde{\rho}^{b/f}_{{\rm Cas}(2)}+\rho^{b/f}_{{\rm Cas(2)div}}
+f^{b/f}_{(2)}\right),
\end{eqnarray}
where $V_1$ is the volume of the 1D noncompact space and the
terms in the last line are the finite part of the Casimir energy density,
the divergent piece, and the temperature dependent contribution, respectively.
As shown in \cite{AFW}, the divergent part of the Casimir
energy is cancelled by the quantum correction to the cosmological
constant, just as in the previous case.  The total free energy density in 2D is
therefore given by,
\begin{eqnarray}\label{second}
  F_2/V_1&=&\sum_{\rm particles}
\left[ \tilde{\rho}^{b/f}_{\rm Cas(2)}+f^{b/f}_{(2)}\right]
+ (2\pi b)^2 \Lambda^{\rm obs}
\nonumber \\
&\equiv& \tilde{\rho}_{{\rm Cas}\Lambda(2)}+f_{(2)}\ ,
\end{eqnarray}
where the formula for $\tilde{\rho}^{b/f}_{\rm Cas(2)}$ is \cite{AFW},
\begin{eqnarray}
&&\!\!\!\!\!\!\!\!\!\tilde{\rho}^{b/f}_{\rm Cas(2)} = \tilde{\rho}^{b/f}_{\rm Cas(2)}(b, m) = \mp \, \frac{d}{(2 \pi b)^2}\nonumber\\
&&\!\!\!\!\!\!\!\times\Bigg[2\left(\tfrac{4}{3}\right)^{1/8}(b \,m)^{3/2}
\sum_{p=1}^\infty \frac{1}{p^{3/2}}K_{3/2}
\left(2\left(\tfrac{3}{4}\right)^{1/4}\pi \,p\, b \,m\right)\nonumber\\
&& \ +
\,\sqrt3\, (b \,m)^2\sum_{p=1}^\infty
\frac{1}{p^{2}}K_2\left(2\left(\tfrac{4}{3}\right)^{1/4}\pi \,p\, b \,m\right)
\nonumber\\
&&\ + \, 4 \left(\tfrac{3}{4}\right)^{1/4} \sum_{n,p = 1}^{\infty}
\frac{1}{p^{3/2}} \left(n^2 +\tfrac{2}{\sqrt3}\,(b \,m)^2\right)^{3/4}
\nonumber\\
&&\ \ \ \ \times(-1)^{p n}\,
K_{3/2}\left(\pi\, p\, \sqrt{3\,n^2 +
2\sqrt3\,(b \,m)^2}\right)\!\Bigg].
\end{eqnarray}
In the second line of equation (\ref{second}), $\tilde{\rho}_{{\rm Cas(2)}\Lambda}$ is the
contribution at zero temperature including the cosmological constant term, while $f_{(2)}$ is
the temperature dependent piece.
The energy density and pressure are calculated
from the total free energy as before,
\begin{align}
\rho_{2} &= \left. \frac{1}{(2 \pi b)^2 V_{1}}
 \frac{\partial \left(\beta F_2\right)}{\partial \beta}\right|_{a,b}\ ,\\
p_{2} &= -\frac{1}{(2 \pi b)^2} \left.
\frac{\partial F_2}{\partial V_{1}} \right|_{b,\beta}\ ,\\
p_b &= -\frac{1}{8 \pi^2 V_{1} b} \left.
\frac{\partial F_2}{\partial b} \right|_{a,\beta}\ .
\end{align}
Einstein's equations take the form,
\begin{eqnarray}
&&\label{oneone} 2 \,\frac{\dot{a}\, \dot{b}}{a\, b} + \frac{\dot{b}^2}{b^2}
=\frac{\rho_{2}}{M_{\rm pl}^2}\ ,\\
&&\label{twotwo} 2 \, \frac{\ddot{b}}{b} + \frac{\dot{b}^2}{b^2}
=-\frac{ p_{2}}{M_{\rm pl}^2}\ ,\\
&&\label{threethree} \frac{\ddot{a}}{a} + \frac{\dot{a}\, \dot{b}}{a\, b}
+ \frac{\ddot{b}}{b} =-\frac{p_b}{M_{\rm pl}^2}\ ,
\end{eqnarray}
and yield,
\begin{align}
\label{boxeq}
\frac{\ddot{b}}{b}+\frac{\dot{a}\, \dot{b}}{a\, b} +\frac{\dot{b}^2}{b^2}
= -\Box{\,\log {b}}
= \frac{1}{2 M_{\rm pl}^2} ( \rho_{2} - p_{2}) \equiv W_2(b)\ .
\end{align}
As in the previous case, the right hand side of (\ref{boxeq}), i.e., $W_2(b)$,
 governs the dynamics of the volume modulus
of our compact space.  Stationary points are given by zeros of
$W_2(b)$. Similar arguments as before show that,
in the cases for which stationary points exist, the one appearing at smaller $b$ is stable, and
the one for larger $b$ is unstable. The $T=0$ results match
the numbers obtained in \cite{AFW}.
 We checked that those zero points also disappear with
increasing temperature.
As an example, figure 2 shows the behavior of $W_2(b)$ for
Dirac neutrinos with normal hierarchy for different temperatures and masses.

At high temperature
the relation $p^T_{2}\simeq\rho^T_{2}$ (where $\rho^T_{2}$ and
$p^T_{2}$ are the temperature dependent parts of the energy density and
pressure, respectively) is satisfied at leading order. One can show at
next-to-leading order that
$\rho^T_{2}-p^T_{2} \sim (b\,T)^2\,T \sim T^3$, where the factor $(b\, T)^2$ again comes from summing Kaluza-Klein modes. Consequently, equation (\ref{boxeq}) becomes,
\begin{eqnarray}
-\Box{\,\log {b}}
\simeq \frac{1}{M_{\rm pl}^2}
\left(\frac{\tilde{\rho}^{b/f}_{{\rm Cas}\Lambda(2)}}{(2\pi b)^2}+\sigma_2\, b^2\,T^3
\right),
\label{eq:2D_b_highTlit}
\end{eqnarray}
where $\sigma_2$ is mass dimension three and contains information on the number of standard model
particles whose mass is smaller than $T$.
Thus $W_{2}(b)$ is an increasing function of
temperature, and for high enough $T$ the zero points
are washed out. As
before, we conclude that there shouldn't have been a ``two-dimensional'' epoch
in the history of our universe yet.

\section{Transitions between spacetimes}
The possibility of transitions between spacetimes of different
dimensionality is certainly interesting to explore
\cite{Sean,Pillado,Graham,Salem,Adamek}.  In our case, as noted in the
previous section, there are no
vacua at high temperature, thus no transition of this type could
have occurred in the early universe. This is in accordance with observation.
Obviously, the current visible universe is still four-dimensional, which also agrees
with the dynamics of the compact dimensions presented in the previous section.

Nevertheless, it seems interesting to investigate the possibility of
a future tunneling of our 4D de Sitter universe to a
lower-dimensional anti de Sitter spacetime, which we have shown to
exist for appropriate neutrino types and masses. Since the current
temperature is $T \simeq 2.7 \rm \ K$ and the universe is cooling
down, we can safely calculate the transition rates using the zero
temperature potential (as was shown in the previous section, for $T
\lesssim 10\rm \ K$ finite temperature effects are negligible).


\subsection{4D $\rightarrow$ 2D transitions}
We first note that a transition of our 4D (false) vacuum to a
2D$\times T^2$ anti de Sitter (true) vacuum \cite{AFW} is only
possible if the geometry of two spatial dimensions of our 4D universe
is that of a ``very large'' torus.  The solutions interpolating
between 4D and 2D$\times T^2$ regions are black holes with toroidal
horizons and such transitions correspond to nucleation of those black
holes in the 4D de Sitter spacetime.  In calculating the transition
rate we follow the usual method outlined in \cite{Sean}\footnote{In
  \cite{Sean} the transition rate from a 6D de Sitter vacuum to a
  4D$\times S^2$ anti de Sitter vacuum was calculated.}.  The
prescription is simple: we take any finite Euclidean solution
connecting one side of the potential barrier with the other, calculate
the instanton action, and use this to estimate the transition rate.

The equations of motion are derived from relations
(\ref{oneone})-(\ref{threethree}) in the limit $T\rightarrow 0$.
The two independent equations reduce to,
\begin{eqnarray}
&&
2\frac{\ddot{b}}{b}+\frac{\dot{b}^2}{b^2}
=\frac{\tilde{\rho}_{{\rm Cas}\Lambda(2)}}{(2 \pi b)^2 M_{\rm pl}^2}
\label{eqm} \ ,
\\ &&
a=C\,\dot{b}  \label{eq:Aeq}   \ ,
\end{eqnarray}
where $C$ is a constant determined by initial conditions.  Analyzing
numerically the dynamics of $b(t)$ we confirmed that the anti de
Sitter vacuum found in \cite{AFW} is indeed stable.

The transition rate is given by,
\begin{eqnarray}\label{gamma}
\Gamma = \Gamma_0 \,e^{-(S_{\rm inst}[b]-S_{dS_4})}\ ,
\end{eqnarray}
where $\Gamma_0$ is a constant, the 4D de Sitter action is \cite{Sean},
\begin{eqnarray}
S_{dS_4}=-24\,\pi^2\frac{M_{\rm pl}^4}{\Lambda^{\rm obs}}
\sim - 10^{122}\ ,
\end{eqnarray}
and $S_{\rm inst}[b]$ is the instanton action.  The equations of motion in
Euclidean space are obtained by flipping the sign of the potential in
equation (\ref{eqm}), as a consequence of the analytic continuation
$t\rightarrow i \tau$.  $S_{\rm inst}[b]$ can be estimated using a finite
numerical (interpolating) solution of the Euclidean equations of
motion connecting both sides of the potential barrier (equivalently,
connecting points on both sides of the Euclidean potential stable
point).  We numerically found finite solutions for the initial
conditions $b(0) = b_0$ and $\dot{b}(0)=0$ for a range of $b_0$
values.  Various finite solutions correspond to nucleating black holes
of different mass in the 4D spacetime.  One of such solutions,
assuming normal hierarchy Dirac neutrinos with the lightest neutrino
mass of $10^{-11} \ \rm GeV$, is shown in figure 3. Using this
solution, the Coleman-de Luccia-like instanton action can be
calculated as \cite{Sean},
\begin{eqnarray}
-S_{\rm inst}[b]
&=& -8\pi^3 M_{\rm pl}^2 \int d\tau
\left[\ddot{a}\,b^2-a\,\dot{b}^2+a \frac{\tilde{\rho}_{{\rm Cas}\Lambda(2)}}
{(2\pi M_{\rm pl})^2}\right] \nonumber\\
&\simeq& 10^{61} \ll -S_{dS_4}\ .
\end{eqnarray}

One can also estimate the transition rate by calculating the
Hawking-Moss-like instanton action which describes $b(t)$ sitting at
the unstable de Sitter vacuum (at $b=b_{dS}$) as,
\begin{eqnarray}
-S_{\rm HM\,inst}=16 \,\pi^3 (M_{\rm pl}b_{dS})^2\simeq
10^{61} \ll -S_{dS_4}\ .
\end{eqnarray}
The Hawking-Moss-like instanton is of the same order as the Coleman-de
Luccia-like instanton obtained via the interpolating solution. Such a
result was expected since the interpolating solution should be equal
to the solution at the top of the de Sitter hill in the limit
$b_0\rightarrow b_{dS}$.  In both cases the transition rate is,
\vspace{-0.1cm}
\begin{eqnarray}
\Gamma \sim e^{S_{dS_4}}\sim e^{-10^{122}}\ .
\end{eqnarray}
This is an extremely small number, so it is clear that these
transitions are highly suppressed.
\begin{figure}[t]
\centerline{\scalebox{1.00}{\includegraphics{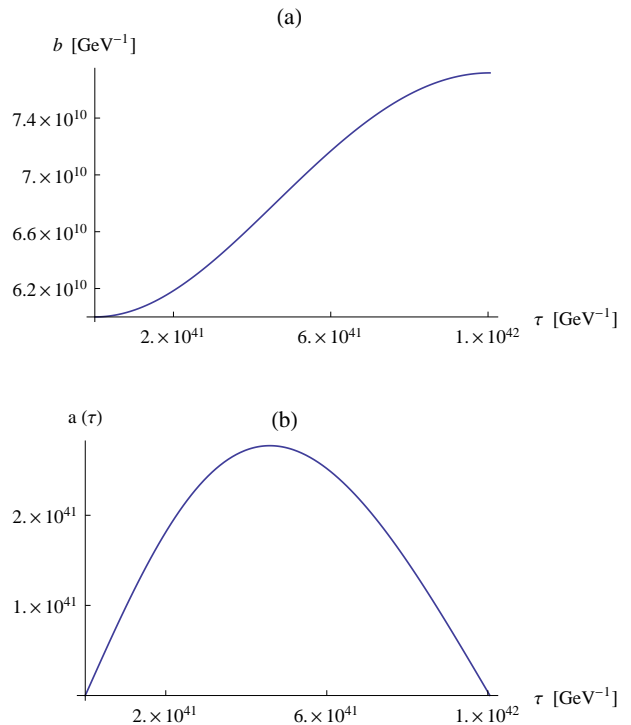}}}
  \caption{\footnotesize{(a) The solution $b(\tau)$ of the Euclidean
      equivalent of equation (\ref{eqm}) assuming normal hierarchy
      Dirac neutrinos with the lightest neutrino mass of $10^{-11} \
      \rm GeV$ and for the initial conditions $b(0) = 6\times 10^{10}
      {\rm \ GeV^{-1}}$ and $\dot{b}(0) = 0$.  (b) The corresponding
      solution $a(\tau)$ of the Euclidean version of equation
      (\ref{eq:Aeq}).\\ \\}}
\end{figure}

\subsection{4D $\rightarrow$ 3D transitions}
We now examine the possibility of transitions from the 4D de Sitter
(false) vacuum to a 3D$\times S^1$ anti de Sitter (true) vacuum
\cite{Arkani}. In the previous subsection we found that the instanton
action corresponding to the transition to a lower-dimensional universe
calculated via the Hawking-Moss-like instanton gives a result of the
same order as the interpolating solution. Here we proceed the other
way around and first estimate the transition rate to a 3D spacetime by
calculating the Hawking-Moss-like instanton. Equations of motion at
zero tempereture can be obtained from relations
(\ref{eq:3D1})-(\ref{eq:3D3}) by taking $T\rightarrow 0$ as in the
previous 2D case.

Now, we have to rewrite and solve these equations in Euclidean space.
We first note that there exists an analytic continuation to Euclidean
space yielding a finite instanton solution only for an open metric
ansatz ($\kappa=-1$) \cite{Sean}.  The equations of motion have the
same form as equations (\ref{eq:3D1})-(\ref{eq:3D3}) after the limit
$T\rightarrow 0$ is taken, but with the substitution
$\tilde{\rho}_{{\rm Cas}\Lambda(3)} \rightarrow - \tilde{\rho}_{{\rm
    Cas}\Lambda(3)} $. In order to calculate the instanton, we take the
solution sitting at the unstable de Sitter vacuum, i.e., at the top of
the $\tilde{\rho}_{{\rm Cas}\Lambda(3)}$ de Sitter hill (at $R=R_0$).
The equations now yield,
\begin{eqnarray}
&& \frac{\dot{a}^2}{a^2}- \frac{1}{a^2}
  +\frac{\tilde{\rho}_{{\rm Cas}\Lambda(3)}(R_0)}{2\pi R_0 M^2_{\rm pl}} =0\ ,
\\
&&\ddot{a}+\frac{\tilde{\rho}_{{\rm Cas}\Lambda(3)}(R_0)}
{2\pi R_0 M^2_{\rm pl}} a=0\ .
\end{eqnarray}
Since $\bar{V}_{(3)}(R_0)>0$, these equations have a solution around
the de Sitter vacuum given by,
\begin{eqnarray}
a(\tau)=\frac{\sin(\omega_0\tau)}{\omega_0}
\label{eq:a_dS3}
\end{eqnarray}
with $\omega_0^2=\tilde{\rho}_{{\rm Cas}\Lambda(3)}(R_0)/(2\pi R_0
M_{\rm pl}^2)$.  The Hawking-Moss-like instanton can now be
approximated by \cite{Sean},
\begin{eqnarray}
  S_{\rm HM\,inst} = - 8 \,\pi^3 \left(
    \frac{2 \pi M_{\rm pl}^6R_0^3}{\tilde{\rho}_{{\rm Cas}\Lambda(3)}({R_0})}
  \right)^{\frac{1}{2}}.
\end{eqnarray}
Using $R_0\simeq 10^{11}~{\rm GeV}^{-1}$ and $V_{(3)}(R_0)\simeq
10^{-35}~{\rm GeV}^{3}$, we obtain $S_{\rm HMinst}\simeq
-10^{92}$. The absolute value of this number is much larger than the 2D case result from the
previous section, however, it is still orders of magnitude smaller
than $-S_{dS_4}$. The transition rate is again extremely suppressed.

We will now search for finite interpolating solutions.  Contrary to
the 2D case, most of the solutions are singular and one has to adopt a
different strategy.  We look for finite solutions around the minimum
of the Euclidean potential, as was presented in \cite{Sean}.  From
equation (\ref{eq:3D_R_T0}), one obtains,
\begin{eqnarray}\label{eq3D}
\ddot{R}+2\frac{\dot{a}}{a}\dot{R}
=
 -\frac{1}{2\pi  M_{\rm pl}^2}
\left(\tfrac{3}{2}\tilde{\rho}_{{\rm Cas}\Lambda (3)}
- \tfrac{1}{2}R\,\tilde{\rho}'_{{\rm Cas}\Lambda (3)}\right)\ .
\end{eqnarray}
After plugging $R=R_0+\delta R$ and using relation (\ref{eq:a_dS3}), equation
(\ref{eq3D}) becomes,
\begin{align}
\!\!\!\!\ddot{\delta R}&+2\omega_0\cot(\omega_0 \tau)\,\dot{\delta R}
\nonumber \\
&\!\!\!\!\!\!\!\!+\!\frac{1}{2\pi  M_{\rm pl}^2}
\left[\tilde{\rho}_{{\rm Cas}\Lambda (3)}^{\prime}(R_0)\!-\!
\tfrac{1}{2}R_0\tilde{\rho}_{{\rm Cas}\Lambda (3)}^{\prime\prime}(R_0)\right]\!
\delta R = 0\ .
\end{align}
This equation has a solution expressed in terms of
Gegenbauer polynomials $C^{(\alpha)}_n$ \cite{Sean},
\begin{eqnarray}
\delta R = K C^{(1)}_n\left(\cos(\omega_0 \tau)\right)\ ,
\end{eqnarray}
where $K \ll R_0$ is a constant, and the index $n$ satisfies the relation,
\begin{eqnarray}
n^2 \!+\!2n\!-\!\frac{1}{2 \pi M^2_{\rm pl}\omega^2_0}
\!\left[\tilde{\rho}_{{\rm Cas}\Lambda (3)}^{\prime}(R_0)\!-\!
\tfrac{1}{2}R_0\tilde{\rho}_{{\rm Cas}\Lambda (3)}^{\prime\prime}(R_0)\right]
=0\,.\nonumber
\end{eqnarray}
In order to have non-singular solutions, the Gegenbauer index $n$ must
be a non-negative integer.  We find that this condition can only be
satisfied for normal hierarchy Dirac neutrinos with the lightest
neutrino mass $\simeq 7.4\times 10^{-12} {\rm \ GeV}$.  A quick check
shows that this mass is associated with a stable 3D de Sitter vacuum,
so the tunneling process is forbidden.

However, as mentioned in \cite{Sean}, it might be possible to obtain
non-singular solutions in another way, i.e., by carefully fine-tuning
the initial conditions rather than looking for a solution around the
unstable de Sitter vacuum. Nevertheless, even if a non-singular
solution exists, it is expected to yield an instanton action of the
same order as the Hawking-Moss-like instanton action. Consequently,
the transition rate is extremely small.

\section{Conclusions}
In this paper, we have investigated the structure of the standard
model coupled to gravity at finite temperature for one- and
two-dimensional compactifications (on a circle and on a torus). For
each case, we have calculated the energy density and pressures from
the free energy. Those quantities enter Einstein's equations, which
govern the evolution of the compactified spacetime at finite
temperature. We have found that stationary points of the equation
of motion for the size of the compact space disappear with
increasing temperature.  The precise temperature at which this
happens depends on the neutrino masses, their hierarchy, and
whether the neutrinos are Dirac or Majorana, but the qualitative
behavior is the same.

For the compactification on a circle we have shown numerically, as an example, that
the stable stationary point disappears at temperatures on the order
of tens of Kelvin, for the lightest neutrino mass of $\sim
10^{-11} \mbox{\rm ~GeV}$, choosing the neutrinos to be
Dirac with normal hierarchy.  With increasing temperature, the
stationary point never reappears.  We have shown analytically that
such behavior is expected in the high temperature limit. The case of
the 2D compactification on a torus is very similar.

Finally, we have calculated the transition rates for tunneling between
vacua of different dimensionality at zero temperature.  Following the
steps outlined in \cite{Sean}, we found Coleman-de Luccia-like
instanton solutions for a tunneling from a 4D de Sitter universe to a
spacetime with two spatial dimensions compactified on a torus.  The
rate for such a process turned out to be extremely suppressed. For the
case of tunneling to a spacetime with one spatial dimension
compactified on a circle, we found only the Hawking-Moss-like instanton
solution, which also yields a negligible transition rate.

Nevertheless, under the condition that the universe has the right
topology, because the transition rates at low temperatures are not
zero, a tunneling process to a lower-dimensional spacetime might be
the ultimate fate of our universe.

\subsection*{Acknowledgment}
The authors would like to express their special thanks
to Mark Wise for inspirational discussions and many extremely helpful comments at all
stages of the work on this paper. We further thank Sean Carroll,
Matthew Johnson and Yu Nakayama for very useful remarks.
The work was
supported in part by the U.S. Department of Energy under contract
No. DE-FG02-92ER40701. K.I. acknowledges the support of the Gordon and
Betty Moore Foundation.\\



\begin{thebibliography}{99}

\bibitem{Arkani}
  N.~Arkani-Hamed, S.~Dubovsky, A.~Nicolis and G.~Villadoro,
  \textit{Quantum horizons of the standard model landscape},
  JHEP {\bf 0706}, 078 (2007)
  [arXiv:hep-th/0703067].

\bibitem{AFW}
  J.~M.~Arnold, B.~Fornal, M.~B.~Wise,
  \textit{Standard model vacua for two-dimensional compactifications},
  JHEP {\bf 1012}, 083 (2010)
  [arXiv:1010.4302 [hep-th]].

\bibitem{Sean}
  S.~M.~Carroll, M.~C.~Johnson and L.~Randall,
  \textit{Dynamical compactification from de Sitter space},
  JHEP {\bf 0911}, 094 (2009)
  [arXiv:0904.3115 [hep-th]].

\bibitem{Kolb}
  F.~S.~Accetta and E.~W.~Kolb,
  \textit{Finite-temperature instability for compactification},
  Phys.\ Rev.\  D {\bf 34}, 1798 (1986).

\bibitem{Kapusta}
  J.~I.~Kapusta and C.~Gale,
  \textit{Finite-temperature field theory},
  Cambridge University Press, Cambridge U.K. (2006).

\bibitem{pdg}
  K.~Nakamura {\it et al.}  [Particle Data Group],
  \textit{Review of particle physics},
  J.\ Phys.\ G {\bf 37}, 075021 (2010).


\bibitem{Pillado}
  J.~J.~Blanco-Pillado, D.~Schwartz-Perlov and A.~Vilenkin,
  \textit{Transdimensional tunneling in the multiverse},
  JCAP {\bf 1005}, 005 (2010)
  [arXiv:0912.4082 [hep-th]].

\bibitem{Graham}
  P.~W.~Graham, R.~Harnik and S.~Rajendran,
  \textit{Observing the dimensionality of our parent vacuum},
  Phys.\ Rev.\  D {\bf 82}, 063524 (2010)
  [arXiv:1003.0236 [hep-th]].

\bibitem{Salem}
  J.~J.~Blanco-Pillado and M.~P.~Salem,
  \textit{Observable effects of anisotropic bubble nucleation},
  JCAP {\bf 1007}, 007 (2010)
  [arXiv:1003.0663 [hep-th]].

\bibitem{Adamek}
  J.~Adamek, D.~Campo and J.~C.~Niemeyer,
  \textit{Anisotropic Kantowski-Sachs universe from gravitational tunneling and its
  observational signatures},
  Phys.\ Rev.\  D {\bf 82}, 086006 (2010)
  [arXiv:1003.3204 [hep-th]].


\end{thebibliography}
\end{document}